\documentclass[usenatbib]{mn2e}
\usepackage{epsf,graphics,graphicx}
\usepackage{amsmath}
\usepackage{amssymb,latexsym,mathrsfs, bm}
\usepackage{color}
\usepackage{float}
\usepackage{natbib}
\usepackage{times}
\usepackage{caption}
\usepackage{subcaption}

\def\eq#1{{Eq.~(\ref{#1})}}

\def\HeII{{{\textrm{He}}~$\rm \scriptstyle II$}}

\title[Equation of state of the intergalactic medium]{Measuring the equation of state of the high$-z$ intergalactic medium using curvature statistics}
\author[Padmanabhan, Srianand and Choudhury]{Hamsa
Padmanabhan$^{1}$\thanks{Electronic address: hamsa@iucaa.ernet.in},
 R. Srianand$^1$\thanks{Electronic address: anand@iucaa.ernet.in},
 T. Roy Choudhury$^2$\thanks{Electronic address:
tirth@ncra.tifr.res.in}\\
$^{1}$ Inter-University Centre for Astronomy and Astrophysics, Pune 411007,
India\\
$^{2}$ National Centre for Radio Astrophysics, Tata Institute of Fundamental Research, Pune 411007, India}

\begin{document}
\date{ }
\maketitle

\begin{abstract}
Using hydrodynamical simulations, we explore the use of the mean and percentiles of the curvature distribution function to recover the equation of state of the high-$z$ ($2 < z < 4$) intergalactic medium (IGM). We find that the mean and percentiles of the absolute curvature distribution exhibit tight correlation with the temperatures measured at respective characteristic overdensities $\bar{\Delta}_i$'s at each redshift. Hence, they provide {complementary} probes of the same underlying temperature-density distribution, and can in principle be used to simultaneously recover both parameters $T_0$ and $\gamma$ of the IGM effective equation of state. We quantify the associated errors in the recovered parameters $T_0$ and $\gamma$ from the intrinsic scatter in the characteristic overdensities and the uncertainties in the curvature measurement. 

\end{abstract}



\section{Introduction}
The Lyman-$\alpha$ absorption lines seen in the spectra of high-redshift quasars arise predominantly from the low to moderate overdensity  intergalactic medium \citep[IGM;][]{cen,zhang,petitjean1995, hernquist, miralda}. The balance between photoionization heating and adiabatic cooling in the IGM  leads to a mean temperature-density relation that is well-approximated \citep{huignedin} by a power-law, $T = T_0 \Delta^{\gamma - 1}$ for overdensities $\Delta \leq 10$. The parameters of this ``equation of state'' depend upon the reionization history of the universe and their predicted values at different epochs vary in different reionization scenarios, so long as either hydrogen or helium reionization happened relatively recently. Otherwise, the temperature-density relation is predominantly set by the shape of the ionizing ultraviolet background and the adiabatic expansion or collapse of large-scale structure \citep{huignedin}. Non-radiative mechanisms of energy injection like, e.g. blazar heating \citep{chang2012, puchwein2012} may also distort the $T-\Delta$ relation. 
Hence, it is important to measure the temperature-density relation at various redshifts in order to constrain the epoch and evolution of reionization and the properties of the ionizing and heating sources.

Detailed studies of different properties of the IGM have contributed to understanding the thermal history of the IGM, by using observed data together with the results of numerical simulations. The methods include (a) using the Doppler $b$-parameter-column density ($b-N$) or $b$-distribution of the Lyman-$\alpha$ absorption lines originating from the IGM \citep{haehnelt1998,schaye1999,schaye2000, mcdonald2001,tirthanandtp1, ricotti2000a,bryan2000, bolton10, bolton12}, (b) using the small-scale power spectrum of the Lyman-$\alpha$ forest \citep{theuns2000a,zaldarriaga2001}, and
(c) using the wavelet decomposition of the Lyman-$\alpha$ lines \citep{theuns2000,zaldarriaga2002,theuns2002, theuns2002a,lidz2010}. The latter two methods do not require Voigt profile decomposition of the spectral lines. Some evidence for a peak in the intergalactic medium temperature around $z \sim 3$, together with a nearly isothermal profile (i.e. $\gamma = 1$) signifying the end of \HeII\ reionization, has been reported \citep{schaye2000, theuns2002a, lidz2010}, however, e.g. \citet{mcdonald2001} do not find evidence supporting this claim.
Interestingly, the parameters ($T_0$ and $\gamma$) derived using the above methods typically have large uncertainties ($\gtrsim$ 30\%).

Recently, the temperature of the IGM over redshifts 4.5 to $\sim 2.8$ has been measured to a high precision ($\lesssim 10$\%) by \citet{becker11} \citep[and extended upto $z \sim 1.5$ by][]{boera2014} using the curvature statistic, which also avoids the fitting of individual spectral lines and can be used to detect the additional heating effect in high-redshift quasar near-zones \citep[][ hereafter Paper I]{hp2014}. The curvature is normally sensitive to both $T_0$ and $\gamma$, however, it is found that at a characteristic overdensity $\bar{\Delta}$, it becomes more sensitive to $T(\bar{\Delta})$ and fairly independent of $\gamma$.  The value of $\bar{\Delta}$ increases with decreasing redshifts, going towards non-linear densities at lower redshifts. The actual value of $\bar{\Delta}$ at each redshift also depends on the effective optical depth in the simulations used to calibrate the observations \citep{boera2014}. This procedure leads to significantly smaller uncertainties in the determination of $T(\bar{\Delta})$ compared to the typical errors in $T_0$ derived using the other methods discussed above. The results indicate a steady rise in $T (\bar{\Delta})$ across  $z \sim 4$ to $z \sim 2$, consistent with additional heat input either from the prolonged reionization of \HeII\ or non-radiative energy injection from blazar heating, etc. A significant contribution to the rise in $T (\bar{\Delta})$ comes from the increase of $\bar{\Delta}$ itself with decreasing redshifts. However, in order to obtain $T_0$ from $T(\bar{\Delta})$, one has to assume the value of $\gamma$ \citep[assumed to be in the range 1.3 - 1.5 in][]{becker11}. Thus, any uncertainty in $\gamma$ will propagate into the uncertainty in the derived value of $T_0$. 

In this letter, we use hydrodynamical simulations to understand the specific regions of the Lyman-$\alpha$ forest that contribute to $\bar{\Delta}$ and their significance for measuring the temperature of the IGM using the curvature statistic.  As the curvature distribution is non-Gaussian, we explore the results on using the percentiles of the distribution together with the mean. We find that this leads to simultaneous constraints on $T_0$ and $\gamma$, thus breaking the degeneracy introduced by $T(\bar{\Delta})$. We propagate the uncertainties in the curvature measurement to associate a resultant error with the recovered values of $T_0$ and $\gamma$. Throughout this letter, we use the cosmological parameters $\Omega_m =
0.26$, $\Omega_{\Lambda} = 0.74$, $\Omega_b h^2 = 0.024$, $h = 0.72$, $\sigma_8
= 0.85$, and $n_s = 0.95$, which are consistent with the third-year Wilkinson Microwave Anisotropy Probe (WMAP) and Lyman-$\alpha$ forest data \citep{seljak2006,viel2006}. The helium fraction by mass is assumed to be 0.24
\citep{oliveskillman}.
\section{IGM simulations and curvature statistics}
\label{sec:simul}
 We perform smoothed-particle hydrodynamical (SPH) simulations using {\sc{gadget-2}} \citep{gadget2} with $512^3$ each of dark matter and baryonic particles in a periodic box of length $10 h^{-1}$ comoving Mpc, which corresponds to redshift intervals of $\Delta z \sim 0.0092, 0.014$ and 0.02 at redshifts 2, 3 and 4 respectively. The ionization correction is applied using the equation of state to assign temperatures to overdensities at each pixel\footnote{This is a simplification at high overdensities, however, we return to the issue of deviation from the power law equation of state at high densities in Sec. \ref{sec:errors}.} and the updated values of background photoionization rates measured by \citet{becker2013} ($ \Gamma_{\rm HI} \times 10^{12} = $ 1.035, 0.789, 0.847 $s^{-1}$ at redshifts 2, 3 and 4 respectively). Simulated spectra (with a spectral sampling of 2.65 km/s per pixel) are generated at each of these redshifts for 100 such randomly drawn lines-of-sight, for 30 different equations of state with $T_0$ ranging from 5000 to 25000 K in steps of 5000 K, and $\gamma$ from 1.1 to 1.6 in steps of 0.1.  In each case, the mean and percentiles of the absolute curvature distribution for all the pixels having normalized transmitted fluxes in the range $0.1 \leq$ flux $\leq 0.9$ are computed.  \footnote{To start with, we use simulated spectra without noise and without applying extra smoothing to the spectrum for exploration of  the different aspects, however we include the effects of noise in the spectra later in Sec. \ref{sec:errors}.}

\begin{figure}
 \begin{center}
 \vskip-0.2in
  \includegraphics[scale = 0.35]{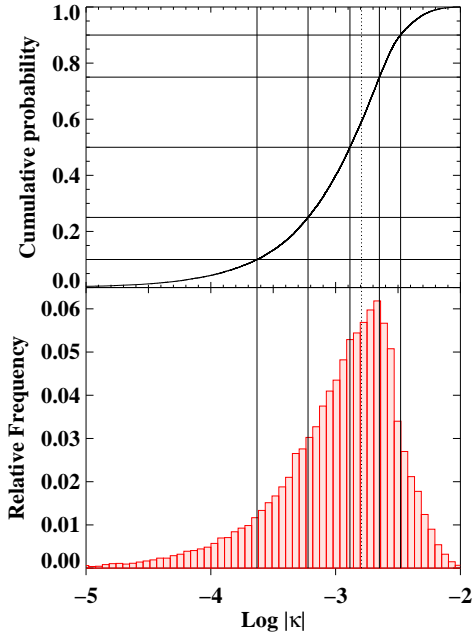}
 \end{center}
\caption{The cumulative probability distribution (top) of $|\kappa|$ and the histogram (bottom) of log $|\kappa|$, for fixed $T_0 = 10000$ K and $\gamma = 1.3$, with 100 lines-of-sight. The percentile values, $D_1, Q_1, Q_2, Q_3, D_9$ are indicated by vertical solid lines and the mean by the vertical dotted line.}
\label{fig:curvdist}
\end{figure}

 The distribution of absolute curvature $|\kappa|$ for the equation of state with  $T_0$ = 10000 K and $\gamma = 1.3$ at redshift 3 is plotted in Fig. \ref{fig:curvdist}. It is clear from the top panel that the curvature distribution is skewed, the mean value is always higher than the median. The mean curvature was shown to follow a tight relationship with the gas temperature at a characteristic overdensity \citep{becker11, boera2014}.  In the present work, we explore the possibility of using the inferred characteristic overdensities at different percentiles of the curvature distribution to simultaneously constrain $T_0$ and $\gamma$.

\begin{figure}
 \includegraphics[angle = 90, scale = 0.33]{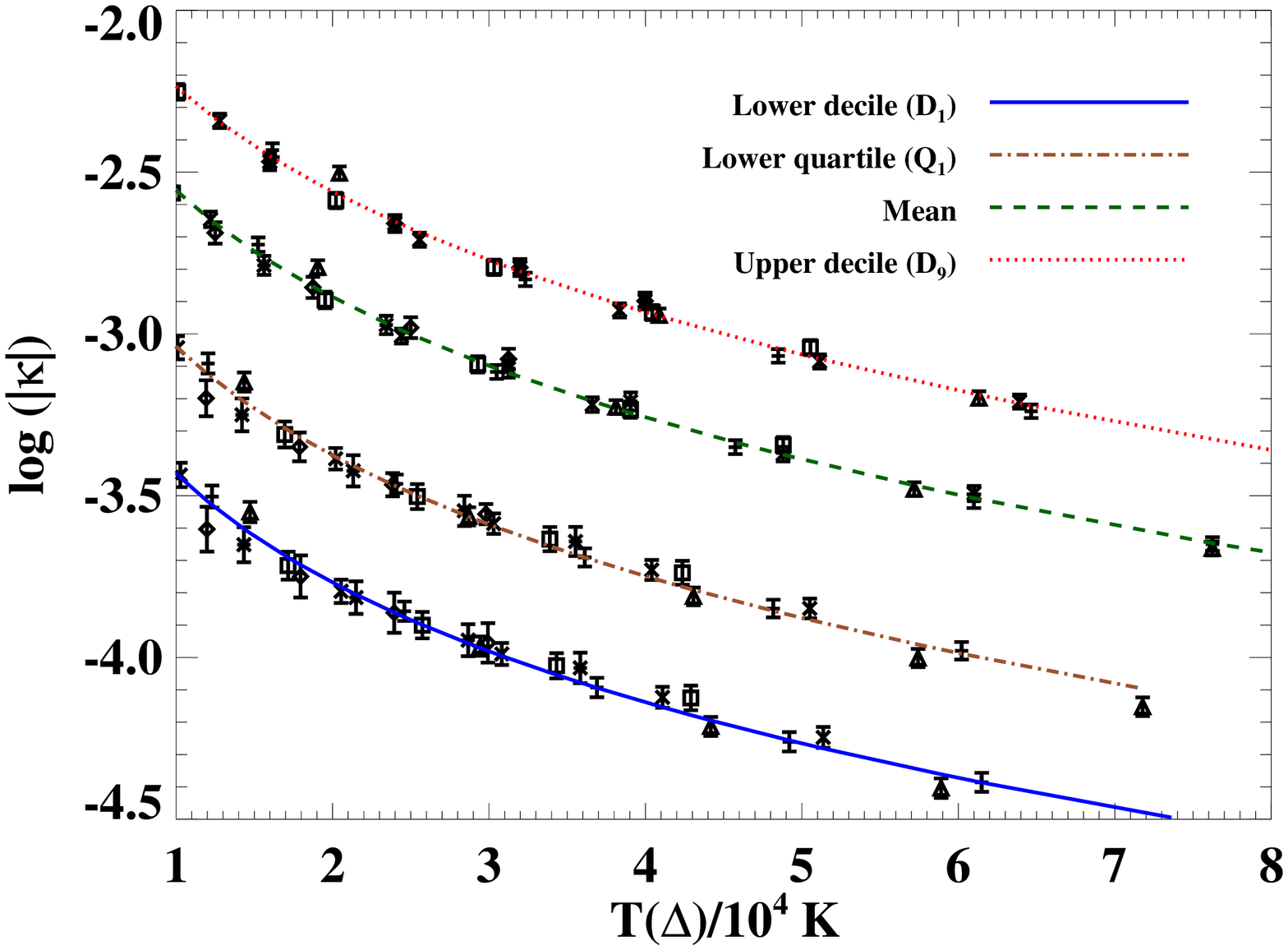}   \includegraphics[angle = 90, scale = 0.33]{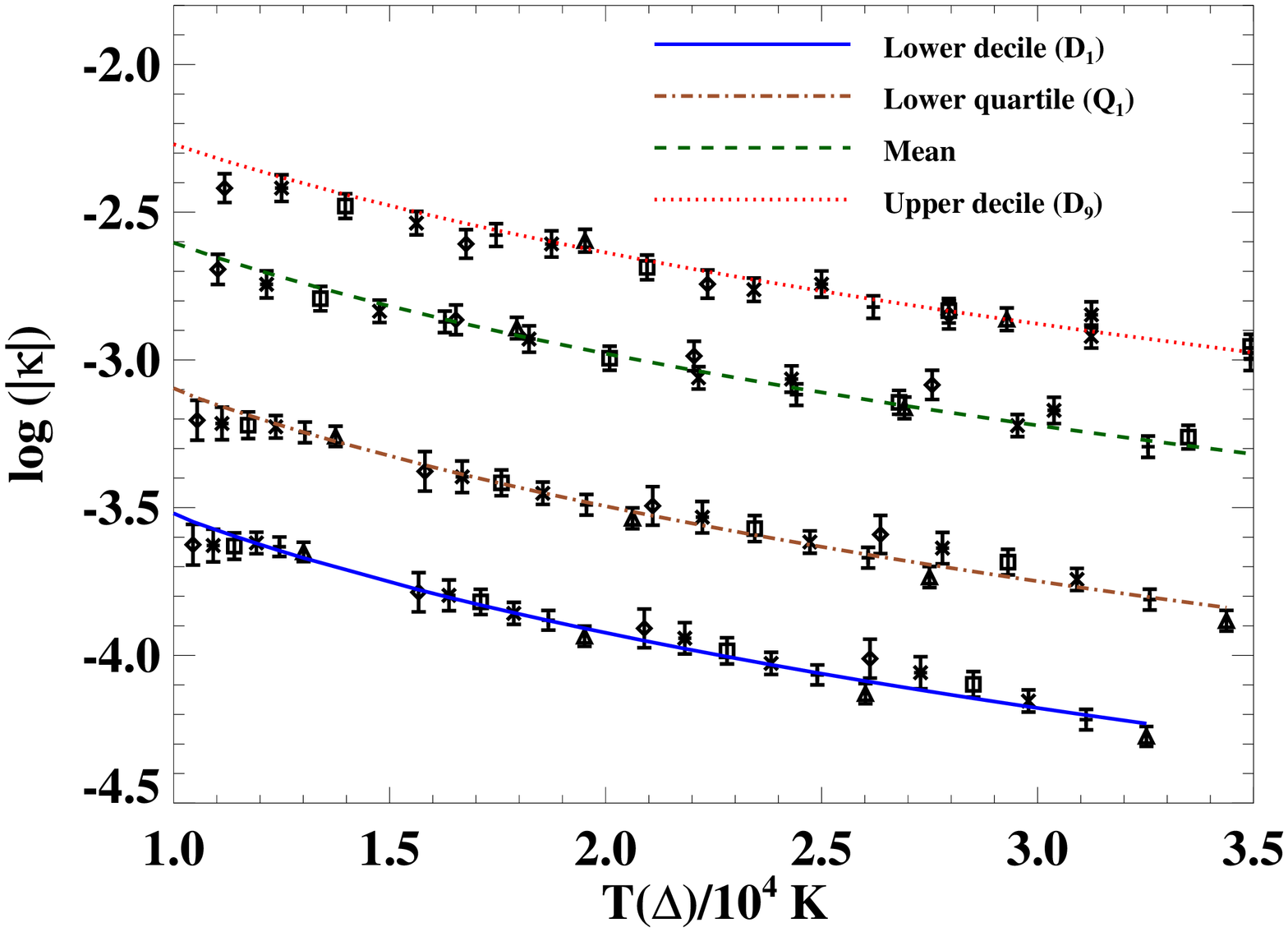} 
 \caption{{Mean and percentiles $D_1, Q_1$ and $D_9$} of the absolute curvature at redshift 2 (top) and redshift 3 (bottom) versus the temperature at the characteristic overdensity. The different colors and linestyles indicate the best-fit curves for the different percentiles. The different symbols indicate the values of $\gamma$ : 1.1 (diamonds), 1.2 (asterisks), 1.3 (squares), 1.4 (crosses), 1.5 (plus signs) and 1.6 (triangles).}
 \label{fig:red3mm}
\end{figure}

At any given $z$, for each of the input models, we find the mean $|\kappa|$ of the simulated spectra. Then, for a given value of $\Delta$, we find $T(\Delta)$ for each model using the assumed $T_0$ and $\gamma$.
We plot the values of $T(\Delta)$ versus $\log\langle|\kappa|\rangle$ for all the input models, and fit the relation using a power law fit with the free parameters $A$ and $\alpha$:
\begin{equation}
 \log \langle|\kappa|\rangle = - \left(\frac{T(\Delta)}{A}\right)^{1/\alpha} 
 \label{powerfit}
\end{equation}
We then vary the value of $\Delta$ in \eq{powerfit}, and find the value of $\Delta$ at which the fit (varying $A$ and $\alpha$) leads to the minimum $\chi^2$. The value of $\Delta$ thus obtained is denoted by $\bar{\Delta}_1$, and defined as the ``characteristic overdensity'' associated with the mean curvature. A similar procedure is applied to find the characteristic overdensities associated with the percentiles $D_1, Q_1, Q_2, Q_3, D_9$ ({the lower (first) decile, lower (first) quartile, median (second quartile), upper (third) quartile and upper (ninth) decile respectively}) of the curvature distribution. Thus, at each redshift, the six characteristic overdensities,  $\bar{\Delta}_i$, $i = 1$ to 6 for the mean and the five percentiles ($D_1, Q_1, Q_2, Q_3, D_9$) respectively are obtained, with corresponding best-fit parameters $A_i$ and $\alpha_i$. Table \ref{table:fit} indicates the values of $A_i$ and $\alpha_i$ for each redshift along with the values of $\bar{\Delta}_i$. 
The best-fit curves for the mean and the percentiles $D_1, Q_1$ and $D_9$ of the curvature are plotted at redshifts 2 and 3 in Fig. \ref{fig:red3mm}, which shows that the mean and percentiles of the absolute curvature follow tight relations with  $T(\bar{\Delta}_i$) at each redshift.\footnote{A slight flattening of the power law fit at very low $T(\Delta) (< 10000 $ K) arises due to a larger fraction of lines being near the saturated regime of the curve of growth. This decreases the sensitivity of the measured curvature to changes in $\gamma$.} 
 We notice that the relationship shown in Fig. \ref{fig:red3mm} is independent of $\gamma$, even when we use  the other percentiles $Q_2, Q_3$ of the curvature.  
 
The scatter in the value of $\Delta$ may be quantified by using the $\chi^2$ of the power law fit. 
  Since we do not have estimates on the systematic errors in $|\kappa|$, we scale the statistical errors in $|\kappa|$ such that the reduced $\chi^2$ at the minimum has the value unity. With this rescaling, we estimate the $1\sigma$ range in the value of the characteristic overdensity, which is found to be $\sigma(\log(\Delta)) \simeq 0.10$. This error is of the same order as that (0.15) estimated by \citet{schaye2000} using the minimization of the scatter in the $\log b_{\Delta} - \log T(\Delta)$ distribution (where $b_{\Delta}$ is the Doppler $b$-parameter corresponding to the density contrast $\Delta$) at the optimal $\Delta$.

\begin{table}
\begin{center}
    \begin{tabular}{ |c | c | c | c  |}
    \hline
      $z$  & $A_i$ & $\alpha_i$ & $\bar{\Delta}_i$  \\ \hline   
   2 &   [46.44, 1.10, 6.54,  &   [5.72, 7.39, 6.60, &  [9.30, 6.05, 5.80, \\
     &   29.07, 83.83, 164.99] &  5.88, 5.41, 5.10] &    6.45, 8.55, 10.45] \\ 
   3 &   [70.57, 2.97, 14.39,  &  [5.17, 6.44, 5.78, &   [2.65, 1.55, 1.70,  \\ 
     &   48.51, 90.33, 224.32] & 5.29, 5.17, 4.63] &   2.05,  2.50, 3.15] \\
   4 &   [73.90, 2.37, 12.71,   &   [5.08,6.67,5.92,   &  [1.55, 1.25, 1.30,  \\ 
     &    47.65, 121.77,253.05] & 5.28, 4.79,4.43] &   1.35, 1.40, 1.60] \\  \hline                                                          
    \end{tabular}
\end{center}
\caption{The best fit parameters $A_i$ and  $\alpha_i$,  $i = 1$ to 6 for the mean and the five percentiles ($D_1, Q_1, Q_2, Q_3, D_9$) respectively, of \eq{powerfit} at three different redshifts. Corresponding values of $\bar{\Delta}_i$ are quoted in the last column. The relative errors on $\bar{\Delta}_i$ are of the order of 0.10.}
 \label{table:fit}
\end{table}

\section{Recovery of temperature}
\label{sec:recovery}
In the present section, we explore the use of the mean and the percentiles of the curvature distribution to recover the parameters of the IGM effective equation of state. The temperature-density relation may be reasonably well-approximated by a power law. We note that the mean and the percentiles of the curvature distribution are typically sensitive to different characteristic overdensities $\bar{\Delta}_i$. In addition, the mean and percentiles are related to specific temperatures $T_i \equiv T(\bar{\Delta}_i)$, through calibration curves analogous to Fig. \ref{fig:red3mm}. Hence, they provide {complementary} probes of the same underlying temperature-density relation. We may, therefore, conjecture using them together to recover both the parameters $T_0$ and $\gamma$ of the equation of state.

The characteristic overdensities $\bar{\Delta}_i$ {for the mean and the $D_1, Q_1, Q_2, Q_3, D_9$ percentiles} of the curvature are provided in Table \ref{table:fit}. From the curvature distribution, we can recover the temperatures $T_i$ associated with the mean and these percentiles.  A power law fit to the recovered temperatures $T_i$ and $\bar{\Delta}_i$ then provides the best-fitting values of $T_0$ and $\gamma$. These can then be compared to the original model.

Applying this procedure at redshift 3 to the spectra drawn from the simulation with $T_0 = 10000 $ K, $\gamma = 1.3$, we recover the best-fit parameters $T_0 = 10399$ K and $\gamma = 1.325$ from a power-law fit to all the percentiles. Hence, the model parameters are recovered to an accuracy of 2-4\%.
This is however an ideal case, since the effects of noise and scatter in the temperature-density relation have to be folded into the analysis in the case of realistic spectra, which we describe in the following section.

\begin{figure*}
 \begin{center}
 \includegraphics[scale = 0.5]{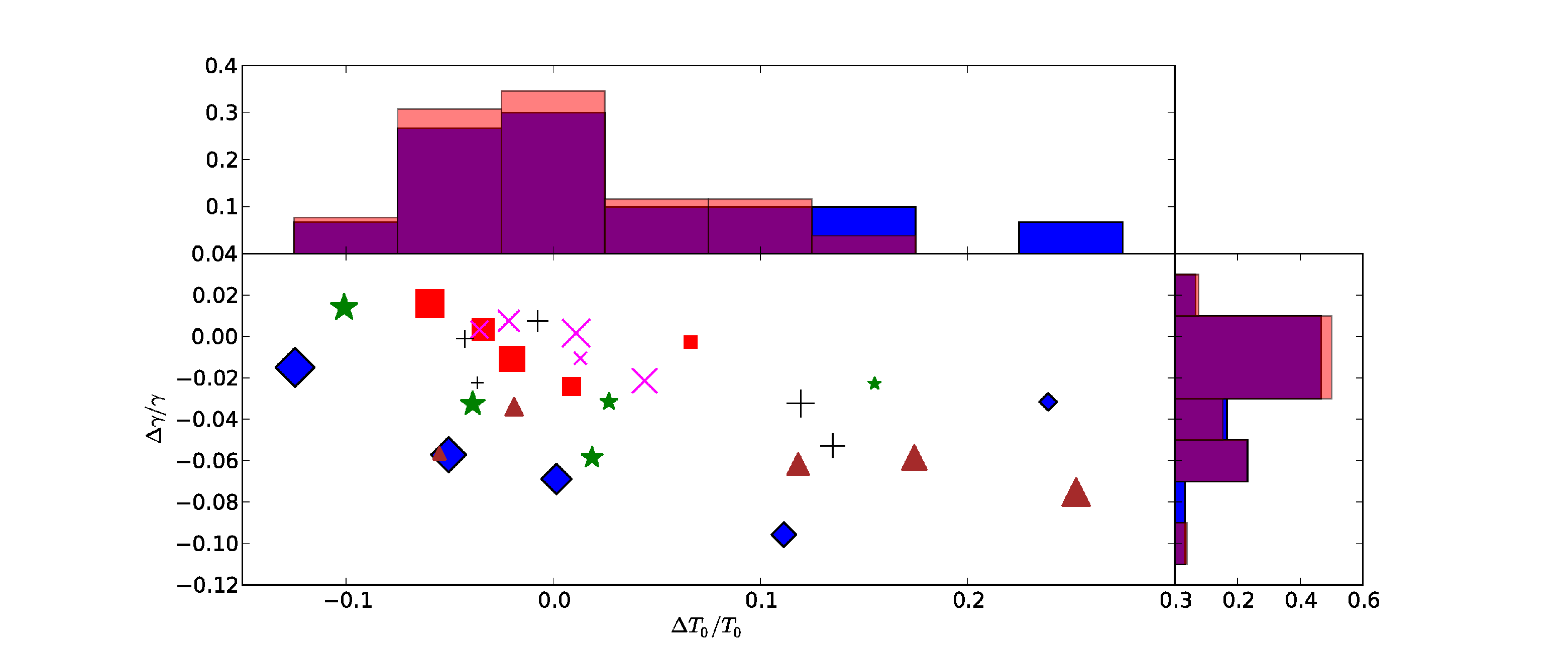} 
 \caption{Distribution of the recovery of $T_0$ [$\Delta T_0/T_0 \equiv (T_{0,\rm recovered} - T_{0, \rm actual})/T_{0, \rm actual}$] and  $\gamma$ [$\Delta \gamma/\gamma \equiv (\gamma_{\rm recovered} - \gamma_{\rm actual})/\gamma_{\rm actual}$] at redshift 3, for the noise added spectra with the deviation from the power-law equation of state. The different models have $\gamma = 1.1$ (diamonds), 1.2 (stars), 1.3 (squares), 1.4 (crosses), 1.5 (plus signs) and  1.6 (triangles), and the size of the symbols is proportional to the value of $T_0 = [5000, 10000, 15000, 20000, 25000]$ K. The recovery uses a best-fitting power law from the mean and all the percentiles of the curvature distribution.  
The histograms of the recovery are shown in the top and right panels. Blue histograms indicate the recovery of $T_0$ and  $\gamma$ for the entire sample. Overplotted are the transparent red histograms which indicate the recovery for the subset of physically motivated models, as described in the text.}
 \label{fig:scatter}
 \end{center}
\end{figure*}

\section{Errors in the recovered parameters}
\label{sec:errors} 
We have seen that the parameters $T_0$ and $\gamma$ of the IGM effective equation of state can be simultaneously recovered by using the mean and percentiles of the curvature distribution. We now explore the effects of addition of noise and incorporating the deviation from the power-law equation of state. 

We begin by estimating the errors in the recovery of $T_0$ due to the uncertainty in the $\bar{\Delta}$ measurement.  We propagate the errors on $\bar{\Delta}$  to the temperature measurement by using the relation \citep{schaye2000}:
 \begin{equation}
  \sigma^2(\ln T_0) =  \sigma^2(\ln T(\Delta)) + \sigma^2(\gamma)[\ln(\Delta)]^2 + (\gamma - 1)^2\sigma^2[\ln(\Delta)] 
 \end{equation} 
with
 \begin{equation}
\sigma^2(\ln T(\Delta)) = \sigma^2(|\kappa|) + \sigma^2_{\rm fit}                                                                  \end{equation} 
 where $\sigma^2_{\rm fit}$ is the scatter of the data points around the fit.
We find  $\sigma (|\kappa|) \sim 0.025$,\footnote{This is the magnitude of the error in log $|\kappa|$ for which the minimum reduced $\chi^2$ has value unity.} and with the conservative estimate of $\sigma_{\rm fit} \sim 0.15$, we obtain $\sigma(\ln T(\Delta)) \sim 0.16$. 
Since $\gamma$ is not known a priori in the realistic case, the error in $\gamma$ is estimated from the recovered value, and this uncertainty in $\gamma$, in turn, propagates to the uncertainty in the recovered value of $T_0$.

It is known \citep[Paper I,][]{becker11, boera2014} that the addition of  noise significantly influences the value of the curvature statistic.  In \citet{becker11}, the noisy spectra are smoothed by using a $b$-spline fitted to the data. In this work, we add noise to have a continuum signal-to-noise ratio of 50 (a typical SNR of e.g. the UVES/HIRES spectra used in IGM studies) and smooth the noisy spectra with a Gaussian filter of width 12 km/s, using a procedure analogous to that described in Paper I.

The temperature of the gas closely follows a power-law equation of state for overdensities $\Delta \leq 10$ \citep{huignedin, puchwein2014} and flattens above $\Delta \sim 10$ \citep{mcdonald2001}. To  take this into account, we simulate noise-added spectra with 10\% Gaussian scatter in the $T-\Delta$ relation, where the equation of state is assumed upto $\Delta = 10$, with a subsequent flattening for $\Delta \gtrsim 10$.\footnote{The flattening of the temperature-density relation for $\Delta \geq 10$ has also been used, for example, in the semi-analytical modelling of the Lyman-$\alpha$ forest power spectrum in \citet{greig2014}.}  
 
Figure \ref{fig:scatter} shows the joint distribution (at redshift 3) of the relative errors in recovery of $T_0$ and $\gamma$ for the noise added spectra incorporating the deviation from the power-law equation of state. The relative error in the recovery  is computed using a best-fitting power law from the mean and all the percentiles of the curvature distribution. The different symbols and sizes indicate the different values of the input parameters $\gamma$ and $T_0$. The blue histograms on the top and right panels indicate the relative errors in recovery for the entire sample. We note that in most realistic scenarios of reionization, the $(T_0,\gamma)$ pairs of (5000 K, 1.1), (5000 K, 1.2), (20000 K, 1.6) and (25000 K, 1.6) are disfavoured, since they correspond to very low or high values of both $T_0$ and $\gamma$. Hence, we plot the recovery for the remaining subset of models as the transparent red histograms, which denote realistic uncertainties for the more physically motivated range in the IGM equations of state.

 \section{Discussion}
We have studied the mean and the percentiles of the curvature distribution of the Lyman-$\alpha$ forest, and their associated characteristic overdensities as probes of the thermal state of the intergalactic medium. 

The characteristic overdensities $\bar{\Delta}_i$ we derive here may be related to the optimal overdensities in the calibration of the $b-T$ relation in \citet{schaye1999, schaye2000, bolton2014}.
It was found that at the optimal overdensity, the $b_{\Delta} - T(\Delta)$ relation becomes independent of $\gamma$, even though e.g. the  $b_{\Delta} - T_{0}$ relation is dependent on $\gamma$. In practice, the value of $\Delta$ is varied and the value corresponding to the minimal scatter in the $b-T$ relation is chosen as the optimal overdensity. Unlike the $b-T$ relation, the $\kappa-T$ relation is nonlinear. However the same trends are found to be valid, i.e. the $\kappa-T$ relation is independent of $\gamma$ even though the $\kappa-T_0$ relation is $\gamma$-dependent \citep{becker11}. 

\begin{figure}
 \begin{center}
  \includegraphics[scale = 0.3, angle = 90]{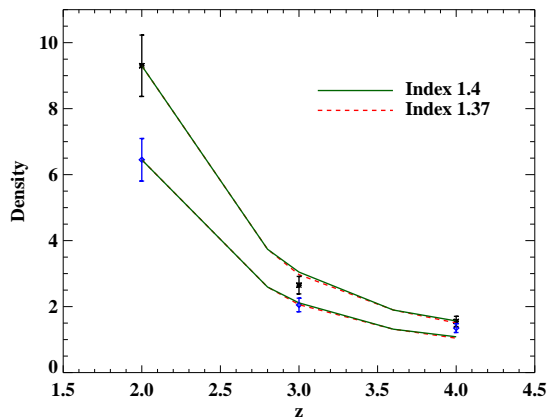}
 \end{center}
\caption{Evolution of the characteristic overdensities associated with the mean ($\bar{\Delta}_1$; black asterisks) and median ($\bar{\Delta}_4$; blue diamonds) curvatures with redshift. Overplotted is the expectation from the evolution of a fixed column density (normalized at redshift 2), following the relation of \citet{schaye2001}. Two different indices of the power law are considered, both of which are found to be a good fit to numerical simulations.}
\label{fig:chardens}
\end{figure}

In \citet{schaye2001}, it is argued that the characteristic size of the absorbers in the Lyman-$\alpha$ forest is of the order of the Jeans' scale. This implies the relation between the neutral hydrogen column density, $N_{\rm HI}$ and the characteristic overdensity $\Delta$ of the absorbers at various redshifts: $N_{\rm HI} \propto \Delta^\alpha \Gamma^{-1} (1+z)^{4.5}$ where $\Gamma$ is the hydrogen photoionization rate. The value of $\alpha$ has been found to be of order 1.37 - 1.43 from the results of numerical simulations \citep{dave1999, schaye2001}. For a fixed column density, the characteristic overdensity is hence expected to scale with redshift as $\Delta \propto (1+z)^{-4.5/\alpha} \Gamma^{1/\alpha}$. We find that the characteristic overdensities roughly follow this scaling (plotted for $\bar{\Delta}_1$ and $\bar{\Delta}_4$ in Fig. \ref{fig:chardens}), assuming the evolution of $\Gamma$ according to \citet{becker2013}.\footnote{At redshift 4, there is a deviation in case of the median, however, note that this range in  $\alpha$ is strictly valid only across redshifts 0-3 \citep{schaye2001}.} Hence, we conclude that these characteristic overdensities trace particular column densities in the Lyman-$\alpha$ forest, which are $N_{\rm HI} \sim 1.6 \times 10^{14}$ cm$^{-2}$ for $\bar{\Delta}_1$ and $9.1 \times 10^{13}$ cm$^{-2}$ for $\bar{\Delta}_4$ (using the relation between the characteristic overdensity and column density). These column densities (as well as those associated with the other percentiles) are in the range $10^{13.8} - 10^{14.3}$ cm$^{-2}$. This is  within the range $10^{12.5} - 10^{14.5}$ cm$^{-2}$ where it is found \citep{schaye1999} that the $b-T$ relation shows minimal scatter. 

We have shown that the mean and percentiles of the curvature act as {complementary} probes of the same underlying $T-\Delta$ relation since they are sensitive to different regions of the Lyman-$\alpha$ forest. Together, they may be used to recover the parameters $T_0$ and $\gamma$ of the IGM effective equation of state. The errors on the recovery of $T_0$ arising from the combination of the scatter in $\bar{\Delta}$, the addition of noise to the spectra and the modifications to the power-law equation of state for realistic models are of the order of $\lesssim 30$\%.  

The above procedure may be applied to the observations of quasar spectra to possibly place interesting constraints on the parameters $T_0$ and $\gamma$ at various redshifts. With the noise as well as the flattening of the equation of state, the final constraints we arrive at are comparable to previous works (e.g. \citet{ricotti2000a}), but one advantage of using the curvature is that Voigt profile line decomposition is not required. The present sample size is consistent with the samples in observations of quasar spectra, e.g. 100-300 sightlines in each redshift interval available from 61 QSO spectra in \citet{becker11}. It is to be noted that increasing the sample size would also considerably reduce the errors. In the realistic case, calibration errors, continuum fitting errors, the impact of Jeans smoothing and the possibility of larger scatter in the equation of state at low densities have to be folded into the analysis, which we plan to carry out in future work.

 \section*{Acknowledgements}
The research of HP is supported by the SPM research grant of the Council for Scientific and Industrial Research (CSIR), India. The hydrodynamical simulations were performed using the Perseus cluster of the IUCAA High Performance Computing Centre. We thank Patrick Petitjean for useful discussions, and the anonymous referee for helpful comments that improved the content and presentation.

\bibliographystyle{mn2e} 
\def\aj{AJ}                   
\def\araa{ARA\&A}             
\def\apj{ApJ}                 
\def\apjl{ApJ}                
\def\apjs{ApJS}               
\def\ao{Appl.Optics}          
\def\apss{Ap\&SS}             
\def\aap{A\&A}                
\def\aapr{A\&A~Rev.}          
\def\aaps{A\&AS}              
\def\azh{AZh}                 
\def\baas{BAAS}
\def\jcap{JCAP}
\def\jrasc{JRASC}             
\def\memras{MmRAS}
\def\na{New Astronomy}
\def\nat{Nature}
\def\mnras{MNRAS}             
\def\pra{Phys.Rev.A}          
\def\prb{Phys.Rev.B}          
\def\prc{Phys.Rev.C}          
\def\prd{Phys.Rev.D}          
\def\prl{Phys.Rev.Lett}       
\def\pasp{PASP}               
\def\pasj{PASJ}
\def\physrep{Phys. Repts.}
\def\qjras{QJRAS}             
\def\skytel{S\&T}             
\def\solphys{Solar~Phys.}     
\def\sovast{Soviet~Ast.}      
\def\ssr{Space~Sci.Rev.}      
\def\zap{ZAp}                 
\let\astap=\aap
\let\apjlett=\apjl
\let\apjsupp=\apjs

\bibliography{mybib}

\end{document}